\documentclass[a4paper,usenames,dvipsnames,11pt]{article}
\usepackage{cite}
\usepackage{tabularx,booktabs,multirow}
\usepackage[makeroom]{cancel}
\usepackage{amssymb}
\usepackage{amsmath}
\usepackage{booktabs, cellspace, hhline}
\usepackage{multirow}
\usepackage{makecell}
\usepackage{graphicx}

\usepackage{xspace}

\usepackage{hyperref}
\hypersetup{colorlinks=true,linkcolor=blue,citecolor=ForestGreen,urlcolor=blue}

\usepackage{xcolor}
\usepackage{setspace}
\usepackage{colortbl}
\newcommand{\LC}{\textrm{LC}}

\newcommand{\FC}{\textrm{FC}}

\usepackage{authblk}
\usepackage{mleftright}

\usepackage[mathscr]{euscript}

\newcommand{\M}{\mathcal{M}}
\newcommand{\A}{\mathcal{A}}
\newcommand{\dd}{\text{d}}

\newcommand{\MG}{\textsc{MadGraph$\_$aMC@NLO}\xspace}

\newcommand{\AmpliCol}{\textsc{AmpliCol}\xspace}

\begin{document}

\title{\textbf{Event generation with exponential scaling in multiplicity using AmpliCol}}

\date{}
\author{
Rikkert Frederix$^{1\,}$\footnote{E-mail:  \texttt{rikkert.frederix@fysik.lu.se}},
Timea Vitos$^{2,3}$\footnote{E-mail:  \texttt{timea.vitos@physics.uu.se}},\\
{\small\it $^{1}$ Department of Physics, Lund University,} 
\\%
{\small\it S\"olvegatan 14A, SE-223 62, Lund, Sweden}\\
{\small\it $^{2}$ Department of Physics and Astronomy, Uppsala University,} 
\\%
{\small\it Box  516,  751 20, Uppsala, Sweden  }\\
{\small\it $^{3}$ Institute for Theoretical Physics, ELTE  E\"otv\"os Lor\'and  University, } 
\\%
{\small\it P\'azm\'any  P\'eter  s\'et\'any  1/A,  H-1117  Budapest,  Hungary}\\
}
\maketitle

\begin{abstract}\noindent
Efficient generation of LHC events is hindered by the rapidly rising cost of evaluating QCD matrix elements with increasing multiplicity. We build on a recently proposed two‑step strategy in which unweighted events are first generated using the leading‑colour (LC) approximation and then reweighted to full‑colour (FC) accuracy, utilising the LC integration efficiency while recovering the exact FC prediction. In this work we extend the method to general Standard Model processes and present \AmpliCol, a standalone implementation designed for LHC collisions. We benchmark multi‑jet, $t\bar{t}$+jets, $ZZ$+jets, and Drell–Yan+jets production, measuring the time required to obtain a fixed number of unweighted events at FC accuracy. Across all processes, the runtime exhibits a stable exponential scaling with multiplicity, far milder than the factorial growth of conventional matrix-element generators. This demonstrates that the \AmpliCol code enables efficient event generation at multiplicities that are otherwise computationally prohibitive.
\end{abstract}
\thispagestyle{empty}
\vfill

\newpage

\begingroup
\hypersetup{linkcolor=black}
\tableofcontents
\endgroup

\section{Introduction}

Accurate simulations of high‑energy collisions are indispensable for both precision measurements and searches for new physics at the LHC~\cite{Campbell:2022qmc}. As experimental analyses increasingly rely on multi‑jet final states and complex event topologies, the demands placed on matrix‑element generators continue to grow. Yet the computational cost of evaluating QCD amplitudes with many external partons remains a fundamental obstacle~\cite{HEPSoftwareFoundation:2017ggl}, even at leading-order (LO) accuracy: the colour algebra proliferates, the integrand becomes sharply structured, and the phase‑space integration slows to a crawl. Even with decades of progress---recursive amplitude construction~\cite{Berends:1987me,Caravaglios:1995cd,Caravaglios:1998yr,Draggiotis:1998gr,Britto:2004ap,Duhr:2006iq}, importance sampling~\cite{Lepage:1977sw,Lepage:2020tgj,vanHameren:2007pt,Kleiss:1994qy,Ohl:1998jn}, improved phase‑space mappings~\cite{Papadopoulos:2000tt,Krauss:2001iv,Maltoni:2002qb,Kilian:2007gr,Gleisberg:2008fv,Alwall:2014hca,Mattelaer:2021xdr,Bothmann:2023siu,Frederix:2024uvy,Sherpa:2024mfk}, stochastic sampling of discrete quantum numbers~\cite{Caravaglios:1998yr,Draggiotis:1998gr,Papadopoulos:2000tt,Mangano:2002ea,Gleisberg:2008fv}, utilising hardware acceleration~\cite{Kanzaki:2010ym,Hagiwara:2010oca,Hagiwara:2013oka,Bothmann:2021nch,Carrazza:2021gpx,Bothmann:2023gew,Cruz-Martinez:2025kwa,Valassi:2021ljk,Valassi:2022dkc,Valassi:2023yud,Hageboeck:2023blb,Valassi:2025xfn}, and machine-learning substitutes~\cite{Bendavid:2017zhk,Klimek:2018mza,Chen:2020nfb,Gao:2020vdv,Bothmann:2020ywa,Gao:2020zvv,Winterhalder:2021ngy,Butter:2022rso,Heimel:2022wyj,Heimel:2023ngj,Heimel:2024wph,Bothmann:2025lwg,Janssen:2025zke,Beccatini:2025tpk}---the scaling of the matrix‑element evaluation still limits the multiplicities that can be reached in practice.

A promising strategy, explored in a recent work~\cite{Frederix:2024uvy}, is to decouple integration efficiency from colour accuracy. Leading‑colour (LC) amplitudes capture the dominant kinematic structure of QCD matrix elements while avoiding the factorial growth of the full‑colour (FC) expansion. This observation motivates a two‑step workflow: generate unweighted events using the LC integrand, and subsequently restore FC accuracy by reweighting each event with the exact colour‑summed matrix element. In earlier work, this approach was shown to yield substantial gains for pure‑QCD processes, with LC amplitudes providing both a tractable integrand and a reweight factor that remains close to unity across the entire phase space.

In this paper we extend this two-step strategy to general Standard Model (SM) processes and present \AmpliCol, a standalone implementation of the LC$\to$FC event‑generation framework. The extension to realistic LHC processes introduces new challenges---notably the treatment of colour‑singlet particles and multiple quark lines, both of which break the one‑to‑one correspondence between colour orderings and efficient phase‑space mappings defined in ref.~\cite{Frederix:2024uvy}. We address these issues through a multi‑channel setup. With these ingredients, \AmpliCol can generate LC unweighted samples efficiently for a broad class of processes and subsequently reweight them to FC accuracy with high secondary unweighting efficiency.

To quantify the performance of this approach, we study four representative LHC processes---multi‑jet, $t\bar{t}+\textrm{jets}$, $ZZ+\textrm{jets}$, and Drell–Yan+jets production---across increasing jet multiplicities. For each process we measure the wall‑time required to obtain $10^5$ unweighted FC events on a single CPU core. Convential matrix-element generators scale factorially with multiplicity, but as we will present later, our results show a clear exponential scaling. We also analyse the scaling of the reweighting step, which grows faster than the LC integration and is expected to become the limiting factor at very high multiplicities. 

The remainder of the paper is organised as follows. Section \ref{sec:method} summarises the two‑step methodology and describes the extensions required for processes with colour singlets. Section \ref{sec:results} presents timing benchmarks and scaling fits for the selected processes. We conclude in section \ref{sec:outlook} and outline future improvements and prospects for integration with existing event‑generation frameworks. Appendix~\ref{appendix} provides additional diagnostics on the secondary unweighting efficiency and effective sample size.

\section{The \AmpliCol approach}\label{sec:method}

The methodology adopted in the \AmpliCol program follows closely that presented in ref.~\cite{Frederix:2024uvy} and is based on a two-step approach for event generation. In the following we present a short summary of its main points. We refer the interested reader to the mentioned reference for further details of the method.

\subsection{Overview of the two-step event generation}

The main hurdle in the computation of tree-level cross sections is the highly complex and peaked matrix element $|\M^2|$ in the integrand,
\begin{align}\label{eq.LC}
\sigma_{\rm FC}\simeq \int |\M|^2\dd\Phi = \int \sum_{i,j} \A_i C_{ij}\A^*_j \dd\Phi,
\end{align}
where we have suppressed the parton density functions, flux factor and additional factors coming from colour and helicity averages. The integral can be rewritten as a double sum over colour-ordered amplitudes $\A_i$ and corresponding colour factors $C_{ij}$~\cite{Paton:1969je,Mangano:1987xk,DelDuca:1999rs,Maltoni:2002mq}, as denoted by the second equality sign, in which the $i,j$ indices label colour orderings of the external QCD particles. The only terms contributing at LC accuracy in this colour sum are those with identical colour order, $i=j$ and so the LC-truncated cross section becomes
\begin{align}\label{eq.LC2}
\sigma_{\rm LC}\simeq \int \sum_i \A_i C_{ii}\A^*_i
\dd\Phi = \sum_{i} C_{ii} \int |\A_i|^2 \dd\Phi.
\end{align}
The two-step event generation proposed in ref.~\cite{Frederix:2024uvy} starts from generating unweighted events using this LC approximation in the first step, treating each (unique) element in the sum as a separate integration channel. In the second step, these LC events are passed through a reweighting algorithm, in which each event weight is multiplied with the reweight factor
\begin{equation}\label{eq.rwfactor}
r^{\LC\to\FC}=\frac{|\M|^2}{\sum_{i} C_{ii}
  |\A_i|^2},
\end{equation}
which is the correction factor for obtaining the FC weight. This correction factor is evaluated using the specific kinematics and helicity of each event, in this way obtaining a FC-accurate (weighted) event sample. Finally, a secondary unweighting is performed in order to obtain equal-weight events.

The main result of ref.~\cite{Frederix:2024uvy} is that this two-step procedure results in efficient event generation for all-QCD (i.e., multi-jet) processes, because the single elements in the sum of eq.~\eqref{eq.LC2} are simple enough that the phase-space parametrisation introduced in ref.~\cite{Byckling:1969luw} captures the structure of the integrand well if the building blocks are ordered according to the colour ordering of $|\A_i|^2$. This results in competitive primary unweighting efficiencies in the LC event generation. Moreover, the reweight factor defined in eq.~\eqref{eq.rwfactor} is close to a constant resulting in secondary unweighting efficiencies in the LC$\to$FC reweighting typically well above 50\%.

For this work, we have extended the \AmpliCol code developed for ref.~\cite{Frederix:2024uvy} to allow for any tree-level process generation in the SM, and focus on presenting results for the runtime to generate $10^5$ unweighted events (at FC accuracy) for standard-candle SM processes relevant to the LHC.

\subsection{Colour singlets}

The main obstacle in the extension of the \AmpliCol program from QCD-only processes to the complete SM is the treatment of colour singlets in the generation of the LC events, i.e., the first step of the two-step procedure. The reason for this difficulty is that the order of the $2\to3$ phase-space blocks introduced in ref.~\cite{Byckling:1969luw} should be in a one-to-one correspondence with a single colour ordering of the LC amplitudes. In that way, the elements of the sum of eq.~\eqref{eq.LC2} can be treated as completely independent channels for which the $2\to3$ phase-space blocks form an efficient phase-space parametrisation, as shown in ref.~\cite{Frederix:2024uvy}. Colour singlets do not directly fit into this picture: a LC approximation does not single out a unique ``ordering'' for colour singlets when there are multiple quark lines and/or multiple colour singlets. Hence, a phase-space parametrisation should be used that takes all the possible orderings for the colour singlets into account. For this we adopt a multi-channel approach with weight optimisation, following ref.~\cite{Kleiss:1994qy}, among all the channels that differ only by the unique positions of the colour singlets in the phase-space ordering.

Similarly to the colour singlets, for processes with more than two quark lines there is also no longer a unique mapping between the colour ordering and the ordering used in the phase-space generation. This is due to the fact that interchanging two quark lines is no longer equivalent to cyclically permuting the colour ordering beyond two-quark-line processes. While we also have implemented a multi-channel procedure to deal with three-quark-line processes, in practice our implementation is not very efficient: including the three-quark line processes seriously deteriorates the results presented below. In the light that contributions with three or more quark lines are subdominant in the computation of cross sections, we refrain from including them, and leave optimising the \AmpliCol code regarding these type of processes for future work.

\section{Results}\label{sec:results}

We use the above method for the event generation of four benchmark LHC processes at center-of-mass energy of $\sqrt{s} = 14$ TeV: 
\begin{eqnarray}
\begin{split}
p p &\rightarrow nj,\\
pp &\rightarrow t\overline{t}+(n-2)j, \\
pp &\rightarrow ZZ+(n-2)j, \\
pp &\rightarrow e^+e^-+(n-1)j,
\end{split}
\end{eqnarray}
with the multiplicity number $n$ varied in the range $n \in [2,7]$. In all processes we use a 5-flavour scheme and set the following cuts on the jets, whenever applicable:
\begin{eqnarray}
p_T(j) > 30 \text{ GeV} \quad,\quad \eta(j) < 6.0 \quad,\quad \Delta R(j_1,j_2) > 0.4.
\end{eqnarray}
In the Drell-Yan processes, a cut on the leptonic invariant mass of $m(e^+e^-)>50$ GeV is used in order to avoid the photon-mediated singularity, and no further cuts on the leptons are placed. We set the renormalisation and factorisation scales equal to the $Z$-boson mass, and use the NNPDF23nlo set~\cite{Ball:2012cx} for the initial-state parton distributions. Wherever applicable, results for single phase-space points and integrated cross sections are cross-checked against the matrix-element generator MadGraph5\_aMC@NLO~\cite{Alwall:2014hca}. We perform all the computations on a single core of an Intel i7-8700K CPU running at 3.70GHz and extract the evaluation time, focusing on how the runtime scales with the multiplicity number $n$.

The total time $t_{\rm tot}$ for the generation of unweighted events at FC accuracy in the two-step approach consists of
\begin{eqnarray}
t_{\rm tot} = t_{\rm int} + t_{\rm rwgt}\, .
\end{eqnarray}
Here, the integration time $t_{\rm int}$ is for the generation of unweighted LC-accurate events. This includes the setup of the integration grids, finding the upper bound of the event weights, and the primary unweighting of the generated events. At most a fraction of 0.1\% of the cross section, the so-called overweight fraction (see e.g.~ref.~\cite{Bothmann:2025lwg}), is truncated to improve the unweighting efficiency, but typically this number is between 0.05 to 0.07\% for the results presented in this work. The time for reweighting plus secondary unweighting $t_{\rm rwgt}$ is required to adjust the weights such that they capture the FC corrections. One must take into consideration the loss of events in the secondary unweighting step and generate the LC event number accordingly, such that the final number of FC events is the required event sample. We consider this fractional loss in the timings of the total time. In the following we compare these various times for the different processes.

\begin{figure}[htb!]
\hspace*{1cm}
\includegraphics[width=\textwidth]{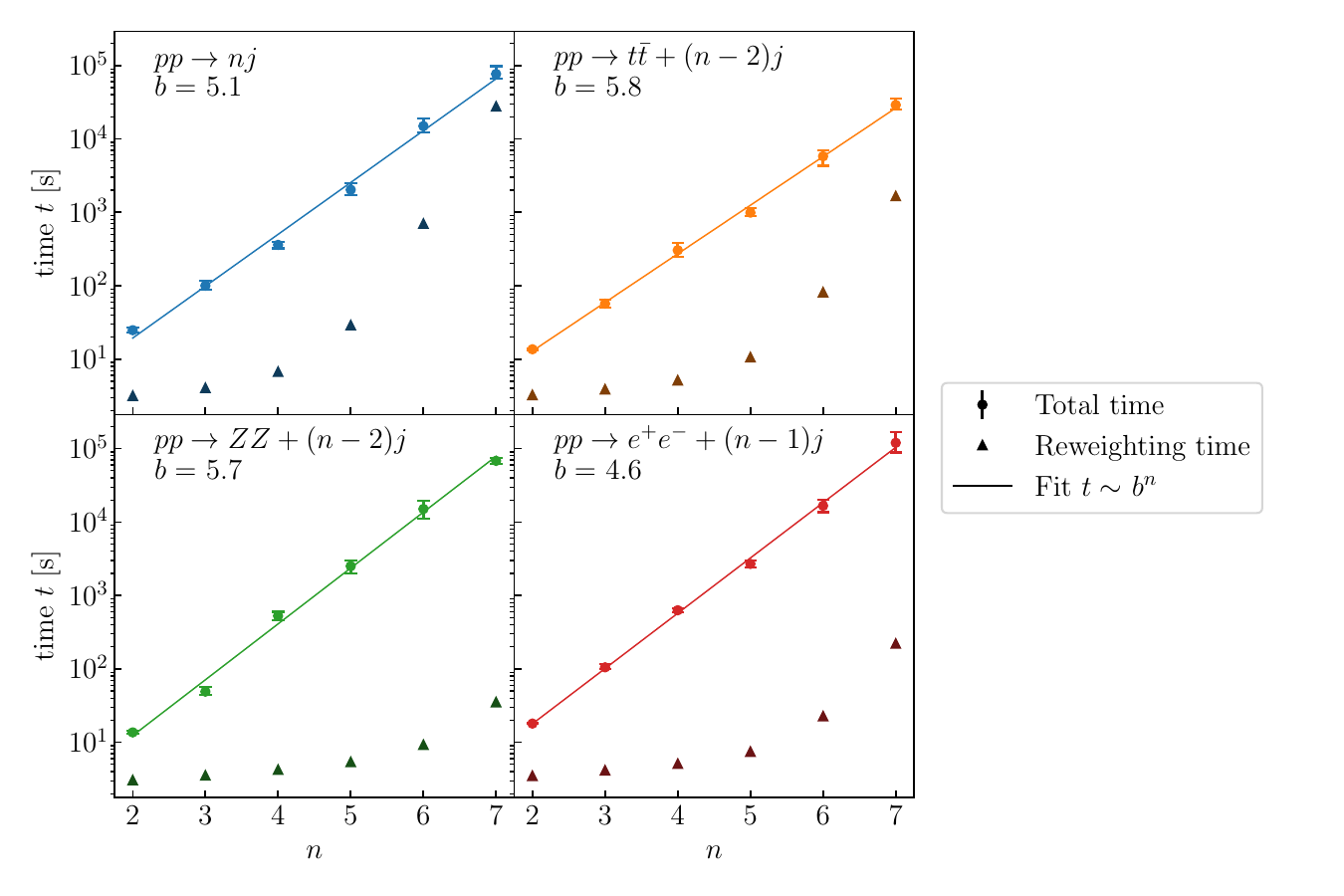}
\caption{Computational time (in seconds) for the four benchmark processes. Shown is the total computation time (circle with error bar) for $10^5$ unweighted events at FC accuracy (including the generation time of the LC events, the reweighting time, and accounting for the secondary unweighting loss), and separately the reweighting time (triangle). An exponential curve is fitted to the total timings and portrayed as a line in the plots, with the base $b$ of the exponential indicated in each of the four sub-plots.}
\label{fig:timings}
\end{figure}

In fig.~\ref{fig:timings} we show the main results of this work: the time to generate $N=10^5$ unweighted events at FC accuracy on a single CPU core. It is shown for the four benchmark processes for varying jet multiplicity. The plot includes the total time $t_{\rm tot}$ required to generate these events (circles with error bars) and also separately the reweighting time $t_{\rm rwgt}$ for each of the processes (triangles). The error bars on the total time are such that they cover the envelope of the total times of 10 separate runs (with different random seeds); the central value is the average. In addition, a fit of an exponential curve is made to the total time with respect to the multiplicity $n$. The fit indicates an exponential growth for all processes with a base of around 5 (with a slight variance between process types, each shown in the figure). 

It is remarkable that an exponential fit covers the total timing results so well. Given that the contribution to the timing from the reweighting is always subdominant---even for $pp \rightarrow 7j$ it is only about a third of the total time, we can conclude that the timing is dominated by the generation of events at LC accuracy. At LC the complexity of the matrix elements scales polynomially ($\sim n^3$) within a given integration channel and for a single helicity configuration, which means that it scales like $\sim 2^n$ due to the fact that \AmpliCol performs an explicit sum over the helicity configurations for each phase-space point. The remaining factor $2.5^n$-$3^n$ comes from the phase-space integration and unweighting: we have explicitly checked that in the \AmpliCol code the matrix elements need to be evaluated about 2.5-3 times more often in order to generate the same number of LC unweighted events when one increases the multiplicity by one. There is also a small increase in the runtime with multiplicity due to the decrease of the secondary unweighting efficiency, i.e.~in the LC$\to$FC reweighting, but this effect is subdominant because it stays approximately within the 60-80\% range as long as there are at least 4 coloured particles in the process, see appendix~\ref{appendix}.

In the current implementation the reweight time $t_{\rm rwgt}$ scales faster-than-exponentially with particle multiplicity, as can be clearly seen from the triangles in fig.~\ref{fig:timings}. Therefore, for large $n$, the time needed for reweighting will become the dominant part of the runtime. From the results in the figure, we expect this to happen when the number of coloured particles in the process (initial plus final state) is equal to ten. If matrix element calculations are needed for such large multiplicities, improvements to the reweight part of \AmpliCol code would be very beneficial. One possibility would be to reweight to next-to-leading colour only~\cite{Frederix:2021wdv,Frederix:2024uvy}; but also ideas along the line of machine learning the reweight factors~\cite{Villadamigo:2025our} and using other colour bases to prevent the beyond-exponential growth of FC matrix elements~\cite{Bolinder:2025gbj} are being investigated.

\section{Discussion and outlook}\label{sec:outlook}

In this work we have presented the program \AmpliCol, which performs the two-step event generation of high-multiplicity tree-level processes, with an exponential increase in computation time up for the multiplicities considered rather than the factorial growth with conventional event generators. We illustrated the computation time for four benchmark processes with various multiplicities. We have shown that with the \AmpliCol code it is possible to generate leading order events for $2\rightarrow n$ processes for $n$ up to $7$ (for $pp\rightarrow \textrm{jets}$, $pp \rightarrow t\bar{t}+\textrm{jets}$ and $pp\rightarrow ZZ+\textrm{jets}$), or $8$ (for $pp\rightarrow e^+e^-+\textrm{jets}$) on a single CPU core with reasonable speed---for the most complicated of these it takes 1-2 days to generate $10^5$ unweighted events.  In the near future, the program will be interfaced with the \MG user-friendly matrix-element generator~\cite{Alwall:2014hca}, improving the automated computation of multi-jet processes, and uncovering multiplicities which are currently out of reach within that framework. Moreover, the \AmpliCol code has a relatively simple structure, and the adaptation to run on multiple cores, or a GPU, should be relatively straight-forward, potentially improving the wall-time for event generation significantly. 

For all the processes presented in this work, the flavour combinations that include three (or more) quark lines have not been included. The reason is not that the code is not capable of including these sub-processes, but rather that the LC event generation efficiency deteriorates significantly when these are included, resulting in much longer runtimes. It is expected that these contributions are more complicated than sub-processes with at most two quark lines: due to various possible orderings of the quark lines at LC, they are no longer in one-to-one correspondence (up to cyclic permutations) with a single, unique ordering used for phase-space generation. We have tried including a simple multi-channel approach, similar to what works for colour singlets where the same problem arises, but without sufficient success. We leave improving the efficiency for processes with multiple quark lines for future work. 

For multiplicities beyond the ones presented in this work, the reweight time to improve the accuracy from LC to FC, will become the dominant part of the runtime. In that case using techniques beyond what is currently used in \AmpliCol will be beneficial. Possibilities that are already being investigated are using machine learning to approximate the LC$\to$FC reweight factors~\cite{Villadamigo:2025our}, and using multiplet bases to tame the factorial growth in the computation of the FC matrix elements~\cite{Bolinder:2025gbj}.

Finally, we expect that efficiency gains observed when using machine learning to improve phase-space sampling, for example within the \textsc{MadNIS} framework~\cite{Heimel:2022wyj,Heimel:2023ngj,Heimel:2024wph}, can also be adopted in the \AmpliCol code. 
\\

\noindent The \AmpliCol program is available from the authors upon request.

\section*{Acknowledgements}

The work of R.F.~is supported by the Swedish Research Council under
project number 202004423.  The work of T.V.~is
supported by the Swedish Research Council under project number
VR:2023-00221.

\appendix

\section{Secondary unweighting efficiency}
\label{appendix}

\begin{figure}[htb!]
\begin{center}
\includegraphics[width=\textwidth]{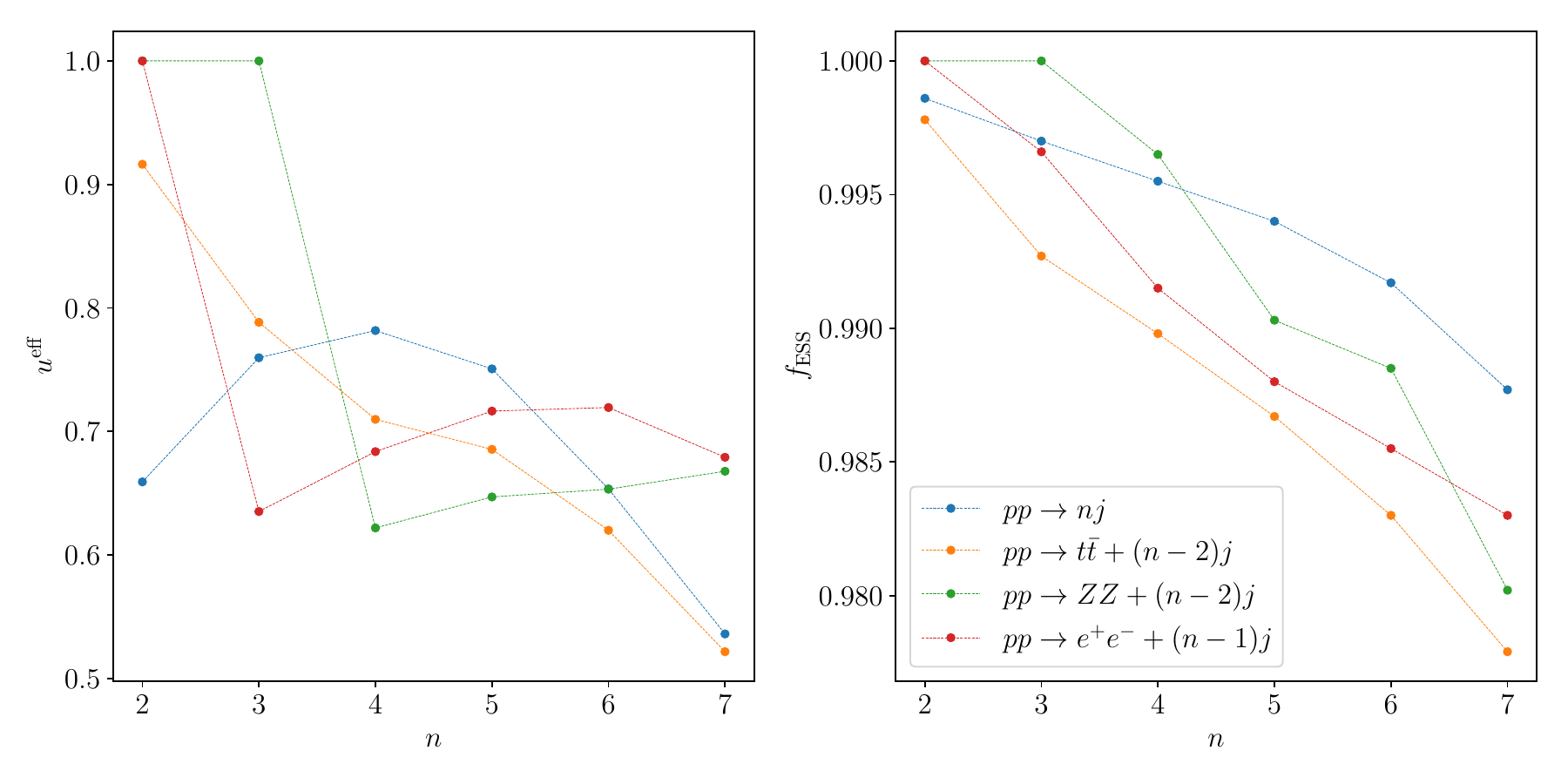}
\caption{The unweighting efficiency $u^{\rm eff}$ for increasing jet multiplicity (left) and effective sample size ratio $f_{\text{ESS}}$ (right) for the four benchmark processes considered.}
\label{fig:ess}
\end{center}
\end{figure}

In our secondary unweighting procedure, i.e.~in the reweighting of the LC events to FC accuracy, we follow the usual procedure of first finding the maximum weight among a sample of events $w_{\rm max}$ and then sampling from the pool of events with the acceptance probability 
\begin{eqnarray}\label{eq:unw_eff}
u^{\rm eff}_i = \frac{w_i}{w_{\rm max}},
\end{eqnarray}
where $w_i$ is proportional to the reweight factor, $r^{\LC\to\FC}$. The fraction of events retained in this procedure is the secondary unweighting efficiency\footnote{Contrary to the primary unweighting, we allow for no overweight fraction since we are dealing with relatively small samples of weighted events.}. In fig.~\ref{fig:ess} (left) we plot the secondary unweighting efficiency for the processes considered. This efficiency is  approximately 60-80\% for all processes, indicating on average a sample size of only 1.3-1.7 times larger at LC accuracy to generate the required number of FC events.

Even though the secondary unweighting efficiency is large, one may consider keeping the weighted events instead. A way of assessing the statistical power of a weighted sample of size $N$ with weights $w_i$ is by evaluating the Kish effective sample size (ESS) ratio~\cite{Kish}, defined by
\begin{eqnarray}
f_{\rm ESS} = \frac{1}{N}\frac{\left(\sum_{i=1}^N w_i\right)^2}{\sum_{i=1}^N w_i^2}.
\end{eqnarray}
This measure evaluates the spread of the weights, reaching a value of 1 for a sample of unweighted events (equal weights). A few outliers from an otherwise almost equal-weight event sample do not impact this effective size greatly, while one outlier will greatly decrease the unweighting efficiency defined in eq.~\eqref{eq:unw_eff}. We show the ESS ratios for the four benchmark processes for varying multiplicity $n$ in the right-hand plot of fig.~\ref{fig:ess}. Even for the highest multiplicities at $n=7$, all of these processes remain at an ESS ratio of beyond 97.5\%, with the multi-jet processes scaling the best with the multiplicity increase. This suggests that keeping the weighted events and thereby reducing the effective sample size by less than 2.5\% significantly outperforms the unweighting of the events, except in cases where post-processing of the events (parton shower, hadronisation, detector simulation, etc.) is particularly time consuming.

\bibliography{Paper}{}
\bibliographystyle{JHEP} 
\end{document}